\def\year{2022}\relax
\documentclass[letterpaper]{article} 
\usepackage{aaai22}  
\usepackage{times}  
\usepackage{helvet}  
\usepackage{courier}  
\usepackage[hyphens]{url}  
\usepackage{graphicx} 
\usepackage{natbib}  
\usepackage{caption} 
\usepackage{subcaption}
\DeclareCaptionStyle{ruled}{labelfont=normalfont,labelsep=colon,strut=off} 
\usepackage{algorithm}
\usepackage{algorithmic}
\usepackage{enumitem}
\usepackage{comment}
\usepackage{multirow}
\usepackage{float}
\usepackage{multirow}
\usepackage{booktabs}
\usepackage{float}
\usepackage{amsmath}
\usepackage{makecell}

\usepackage{newfloat}
\usepackage{listings}
\usepackage{marvosym}
\floatstyle{ruled}
\newfloat{listing}{tb}{lst}{}
\floatname{listing}{Listing}
\title{YES SIR!\\Optimizing Semantic Space of Negatives with Self-Involvement Ranker}
\author{
    Ruizhi Pu\equalcontrib\textsuperscript{1},
    Xinyu Zhang\equalcontrib\textsuperscript{1},
    Ruofei Lai\equalcontrib\textsuperscript{1},
    Zikai Guo\textsuperscript{1},
    Yinxia Zhang\textsuperscript{1},
    Hao Jiang\textsuperscript{1},
    Yongkang Wu\textsuperscript{1},
    Yantao Jia\textsuperscript{1},
    Zhicheng Dou\textsuperscript{2},
    Zhao Cao\textsuperscript{\Letter}\textsuperscript{1}
}
\affiliations{
    \textsuperscript{\rm 1}Distributed and Parallel Software Lab, Huawei\\
    \textsuperscript{\rm 2}Gaoling School of Artificial Intelligence, Renmin University of China\\

    {zhangxinyu35,caozhao1}@huawei.com
}

\usepackage{color}

\begin{document}

\maketitle

\begin{abstract}
Pre-trained model such as BERT has been proved to be an effective tool for dealing with Information Retrieval (IR) problems. 
Due to its inspiring performance, it has been widely used to tackle with real-world IR problems such as document ranking. Recently, researchers have found that selecting ``hard'' rather than ''random'' negative samples would be beneficial for fine-tuning pre-trained models on ranking tasks. However, it remains elusive how to leverage hard negative samples in a principled way. To address aforementioned issues, 
we propose a fine-tuning strategy for document ranking, namely Self-Involvement Ranker (SIR), to \textbf{dynamically} select \textbf{hard} negative samples to construct high-quality semantic space for training a high-quality ranking model. Specifically, SIR consists of sequential \textbf{compressor}s implemented with pre-trained models. Front compressor selects hard negative samples for rear compressor. Moreover, SIR leverages supervisory signal to adaptively adjust semantic space of negative samples. Finally, supervisory signal in rear compressor is computed based on condition probability and thus can control sample dynamic and further enhance the model performance. SIR is a \textbf{lightweight} and \textbf{general} framework for pre-trained models, which simplifies the ranking process in industry practice. We test our proposed solution on MS MARCO with document ranking setting, and the results show that SIR can significantly improve the ranking performance of various pre-trained models. Moreover, our method became \textbf{the new SOTA model anonymously} on MS MARCO Document ranking leaderboard in May 2021.

\end{abstract}

\maketitle

\section{Introduction}
Information Retrieval (IR) which aims to obtain specific information from the Web based on queries has become one of the most prevailing research topics. A key component in an IR system is the ranking model which sorts the documents based on their relevance to a query. Classic ranking algorithms, such as BM25~\cite{robertson1994some}, simply rely on statistics of keyword matching and suffer from the vocabulary mismatching problems. Because of this, many embedding-based neural ranking models have been proposed to overcome this problem and have achieved better performance~\cite{10.1145/2505515.2505665,shen2014learning,wan2016deep}. 
In recent years, Pre-trained Language Models (PLM), which have fundamentally changed many neural language process areas, have also been applied to document ranking. PLM can effectively capture contextual information and semantic information from text~\cite{devlin2019bert,liu2019multitask}, and hence would bring better match between a query and a document. PLM-based document ranking models have achieved significant improvement~\cite{lample2019crosslingual}. 

\begin{figure}[t]
     \centering
    \includegraphics[width=0.45\textwidth]{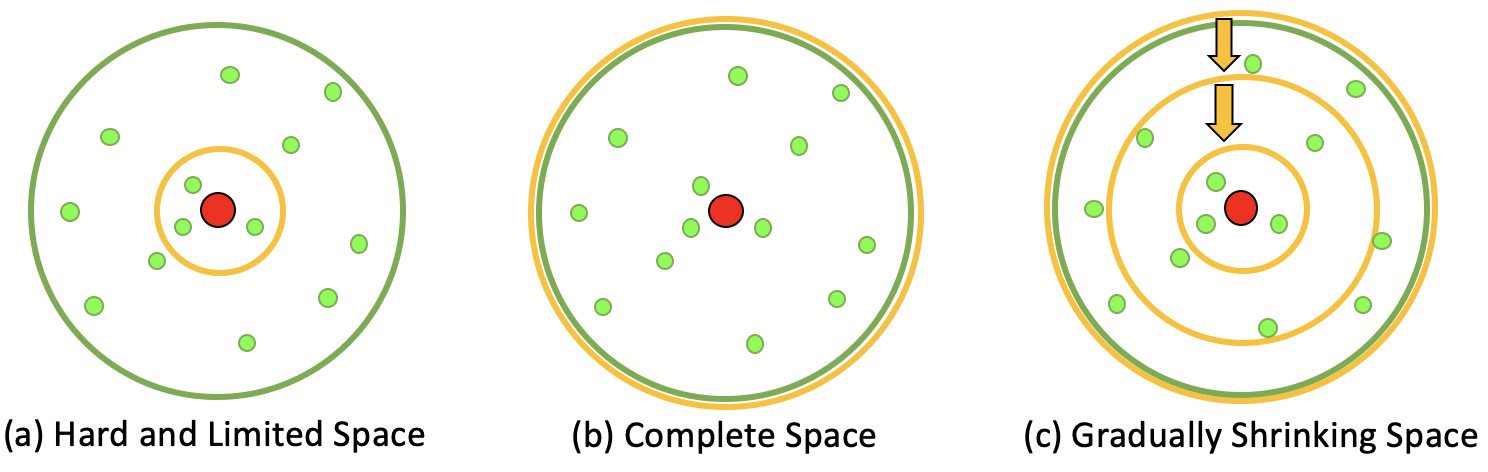}
        \caption{Illustration of three fine-tuning strategies. (a): Select hard samples for training. (b): Select random samples for training. (c): Gradually identify and select hard samples. Here, red points represent positive samples and green points represent negative samples. Green circle represents negative sample space, limited by retrieval stage. Orange circle represents training negative sample space. }
        \label{fig2}
\end{figure}

As the goal of the document ranking is different from other NLP tasks, directly applying existing general PLMs (such as BERT~\cite{devlin2019bert} or GPT~\cite{brown2020language}) to document ranking is not optimal. To solve this problem, researchers proposed to design new IR-oriented objectives or self-supervised tasks for training pre-trained models specially for IR. For example, ~\cite{DBLP:journals/corr/abs-1906-00300, DBLP:journals/corr/abs-2002-03932} proposed Wiki Link Prediction (WLP) task and Inverse Cloze (ICT) task and~\cite{10.1145/3437963.3441777} proposed Representative Words Prediction task (PROP). We refer to them as pre-trained IR models.

Although these pre-trained models have proved to be able to significantly improve ranking quality, it is still under investigation on how to effectively leverage these pre-trained models to train a ranking model. Ranking models are typically trained under Learning to Rank (LTR)~\cite{10.1145/2505515.2505665} framework. Take pair-wise loss based LTR for example, ranking models focus on the relative order between a positive sample and a negative sample, for a given query. When constructing train sets for fine-tuning, a common approach is to use random sampling methods for negative sample selection~\cite{Huang_2020}. But such approach is sub-optimal, because randomly-selected negative samples may be uninformative~\cite{xiong2020approximate,robinson2021contrastive} and the distribution of negative samples are fixed. Recently, some studies~\cite{xiong2020approximate, DBLP:journals/corr/abs-2104-08051, gao2021rethink} showed that a proper selection of negative samples is essential to training a ranking model by fine-tuning PLMs. 
Motivated by human learning process and curriculum learning~\cite{chevalier2018babyai, bengio2009curriculum}, we observe that human tend to learn knowledge from easy to hard. For pairwise ranking models, the difficulty of negative samples can be measured by their relevance to the query, which is reflected as their predicted scores. The model should first distinguish very irrelevant negative samples from positive one and then distinguish confusing negative samples from positive sample.

We show the differences between three fine-tuning strategies in Figure \ref{fig2}. In Figure \ref{fig2}a, only the hardest negative samples are selected for training and the semantic space of negatives is quite limited. In Figure \ref{fig2}b, negative samples are randomly selected and the semantic space is relatively complete, but with unreliable quality. In Figure \ref{fig2}c, negative samples are selected from easy to hard. The semantic space is complete and of high quality.

In this paper, we propose a novel training paradigm for neural ranking models atop pre-trained models, which is called \textbf{S}elf-\textbf{I}nvolvement \textbf{R}anker (\textbf{SIR}). SIR focuses on optimizing semantic space of negative samples when fine-tuning ranking models. To fulfill the goal of gradually increasing learning difficulty, we use multi-level ``compressor'' to select hard negative samples step-by-step. We use the terminology compressor because the number of training samples after each compressor would considerably decrease. Moreover, in order to achieve self-adaptation and adjustment for negative semantic space, we construct an end-to-end derivable framework which can adjust hard negative samples dynamically by supervisory signals. Here, dynamic hard samples means that hard samples would change as model parameter changes. Meanwhile, in this work, we make dynamic under control by using an conditional probability-liked controller to constrain it, which proves to be an effective way to improve the model performance. Finally, to further explore the learning process of hard negative semantic space, we conduct multiple comparison experiments based on SIR, including SIR V1, V2, V3 and V4. And we highlight SIR V3 and V4 in our work due to their outstanding performance.

%

\begin{figure}[!tb]
\centering
\includegraphics[width=0.45\textwidth]{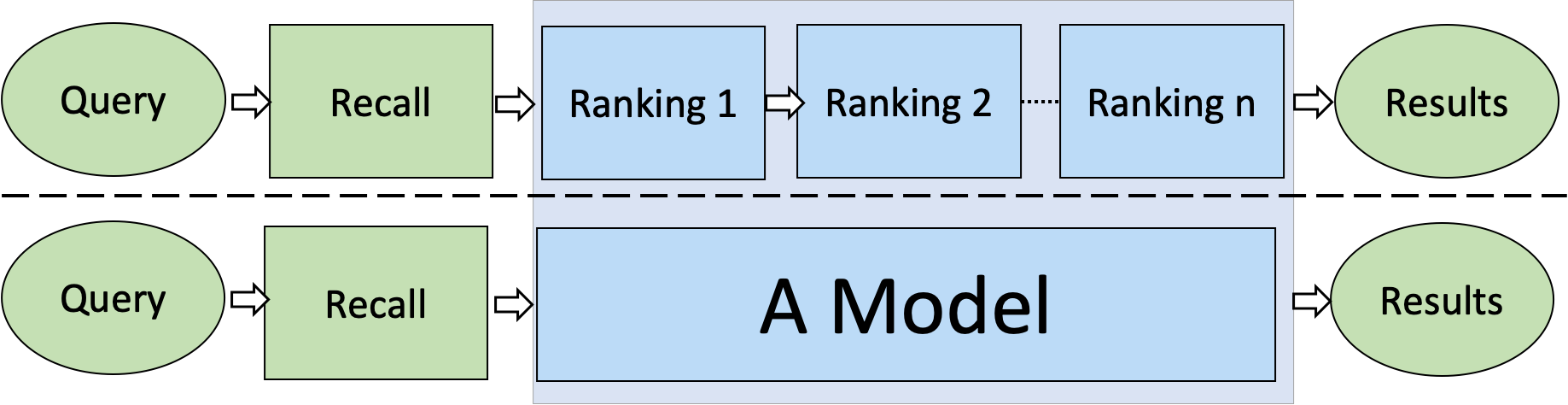}
\caption{Upper: Pipeline of ranking models in industry practice. Lower: Pipeline of SIR ranking models.}
\label{pipeline}
\end{figure}

Besides performance improvement, SIR also simplifies the pipeline of ranking models in industry practice. Ranking models in industry usually have a complex pipeline, which is comprised of several ranking stages~\cite{feng2021revisit}, as shown in Figure \ref{pipeline}. In contrast, out model just uses only ``A Model'' to leverage the whole ranking process, which reveals great simplification for ranking process.

Experimental results show that SIR can significantly enhance the ranking performance of pre-trained language models and pre-trained IR models. Meanwhile, SIR became the new SOTA model anonymously on MS MARCO Document Ranking leaderboard in May 2021.

\textbf{In summary, our contributions are as follows}:

\begin{itemize}
\setlength{\itemsep}{0pt}
\setlength{\parsep}{0pt}
\setlength{\parskip}{0pt}
\item 
We propose a novel training paradigm for neural ranking models atop pre-trained models named SIR, which dynamically selects  negative samples to construct high-quality semantic space. SIR can simplify ranking pipeline in industry practice.
\item
We propose to control sample dynamic via supervisory signal, which is computed based on conditional probability. Such mechanism further improves semantic space.
\item
Experiments on MS MARCO show that SIR significantly enhances both pre-trained language models and pre-trained IR models.
\end{itemize}

\begin{figure*}[t]
     \centering
         \includegraphics[width=\textwidth]{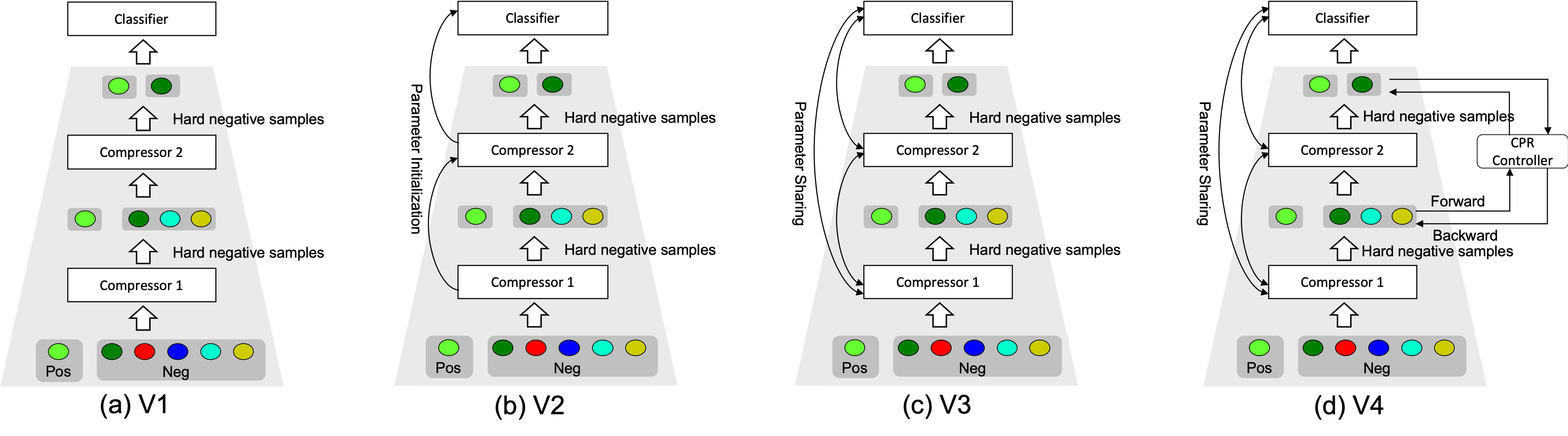}
         \caption{SIR Structure: Left to Right from V1 to V4. Here, "Pos" is the short form of positive samples and "Neg" is the short form of negative samples. "CPR" is the short form of conditional probability.}
         \label{fig3}
\end{figure*}

\section{Related Work} \label{Related}

\subsection{Neural Document Ranking}
Neural document ranking models compute relevance scores for query-document pairs based on neural networks. It is critical for them to learn good representation of queries, documents and ranking functions. 

We can categorize neural ranking models to representation-focused models and interaction-focused models~\cite{trabelsi2021neural}. The representation-focused models extract good feature representation from input data.~\cite{10.1145/2505515.2505665} first utilizes deep neural networks to extracts features from query and documents independently. To capture local context,~\cite{shen2014learning, qiu2015convolutional} utilize convolutional neural networks instead of feed-forward networks. Also, recurrent neural networks are used in~\cite{wan2016deep} to capture richer context. In contrast, the interaction-focused models aim to capture matching signals from interactions of query and document tokens.~\cite{guo2016deep} performs term matching using histogram-based features.~\cite{wan2016match, pang2017deeprank} capture matching signals with gated recurrent units, while~\cite{dai2018convolutional, hui2017pacrr} use convolutional neural networks.

Neural ranking models are typically trained using the Learning to Rank (LTR)~\cite{10.1145/2505515.2505665} framework. LTR framework can be divided to three categories based on the training objectives: pointwise, pairwise and listwise~\cite{liu2011learning}.

\subsection{Pre-trained Models for Document Ranking}

The great success of pre-trained language models inspires researchers to utilize pre-trained models for document ranking. A straightforward way is to use PLM, such as BERT or ELECTRA~\cite{clark2020electra}, to accomplish ranking tasks. 

To further enhance models' capacity on IR, researchers have proposed several pre-trained IR models, whose pre-training tasks are highly relevant to downstream IR tasks. $\text{Transformer}_{ICT}$~\cite{DBLP:journals/corr/abs-1906-00300} designs pre-training tasks to explicitly explores the target context based on the query sentence.
\cite{DBLP:journals/corr/abs-2002-03932} uses random sentence from Wiki as the query to retrieve the related passage from same page or from the hyper-linked page. Ma has explored the representative words prediction~\cite{10.1145/3437963.3441777} for ad-hoc retrieval and later he proposed B-PROP~\cite{DBLP:journals/corr/abs-2104-09791} to free PROP from limitation of classical uni-gram language model.

Another way to enhance pre-trained models is to design fine-tuning strategies. ULMFiT~\cite{DBLP:journals/corr/abs-1801-06146} is the pioneering work which uses several strategies to fine-tune the pre-trained model on the text classifiers. Further research tries to use the fine-grained sample strategies to design tasks, such as~\cite{Huang_2020} utilized random negative sampling for recall tasks while ANCE~\cite{xiong2020approximate} used the top of recalled hard negative samples for training. It is surprising that hard negative samples seem experimentally conductive for downstream tasks.~\cite{DBLP:journals/corr/abs-2104-08051} theoretically explains the effectiveness of hard negative samples and also introduces two strategies for both dense retrieval and reranking.

\section{Methodologies} \label{Methodologies}


In this section, we design 4 different negative sample selection strategies that belong to SIR framework. All of 4 strategies aim to select appropriate negative samples through compressor, which is implemented by pre-trained models. From V1 to V4, the selection of negative samples gradually tends to be adaptive and dynamic. The structure is shown in Figure \ref{fig3} and the implementation of each version will be introduced in the following subsections.

Given one query $q$ and one corpus $D$ (including corresponding positive sample corpus $D^+$ and negative sample corpus $D^-$), traditional train set construction methods like~\cite{10.1145/3437963.3441777} mostly use one query $q$ patching with one positive document sample $d^{+} \in D^{+}$ and one negative document sample $d^{-} \in D^{-}$. Such method is straightforward but information from negative samples, which might be conductive to IR objectives, is ignored. For this reason, we construct one input train set in a contrast manner which combines one positive document sample with $N$ negative samples. We aim to select most informative and critical negative samples for training.

\subsection{Strategy V1: Self-involvement with Hard Samples}

In strategy V1, the trained front compressor will select the top negative samples with higher relevance scores (higher difficulty), and provide them to the rear compressor for training. The process has been shown in Figure \ref{fig3}a.

Supposing that there are $M$ compressors $\mathcal{C}$ (including classifier). The compressor aims to correctly classify positive sample as well as select hard negative samples. As mentioned before, compressors are implemented by some pre-trained models. All compressors have identical network structure in our settings.



We denote parameter of all compressors as $\theta$. For $i$-th compressor $C_i$, we denote its parameter as $\theta_i$, network as $f_i$ and input data as $S_i$. For compressor $C_1$, its input data $S_1$ is a set of queries $Q$ and their corresponding samples. Each query would correspond to one positive sample and $N$ negative samples. We use $|S_i|$ to denote number of negative samples pairing a positive sample for $i$-th compressor, therefore $|S_1|=N$. For query $q$, its training sample $S_i(q)$, would be $(q, d)$. $d= (d^{0}, d^{1}, d^{2},..., d^{|S_i|})$. Here $ d^{0} \in D^{+}_q$ and $d^{1},..., d^{N} \in D^{-}_q$. Note that the positive sample would be fixed at the first index position following with $|S_i|$ negative samples. $i$-th compressor outputs relevance scores $o_i \in R^{1+|S_i|}$ for input samples. $o_i= f_{i}(S_i)$. For $i$-th compressor, we can formulate our objective function on query $q$ as cross entropy loss:

\begin{align}
\begin{split}
        \mathcal{L}(S(q), \theta, i) =  - \log P(o^{0}_i) - \sum_{j=1}^{|S_i|}\log (1- P( o^{j}_i )),    \label{loss}
\end{split}  
\end{align}
where $P$ denotes the probability function.

\begin{align} \label{softmax}
\begin{split}
        P(o^j_i) = \frac{\exp {(o^j_i)}}{\sum_{j=0}^{|S_i|} \exp {(o^j_i)}}.
\end{split}  
\end{align}

We can calculate $\theta_i$ as:
\begin{align}
    \begin{split}
        \theta_i  = \arg\min_{\theta_i} \sum_{q\in Q}\mathcal{L}(S(q), \theta, i).  \label{theta}
    \end{split}
\end{align}

After training the first compressor, the model will evaluate the difficulty of input negative samples and then pass hard ones to the next compressor.
We refer this infer-after-train procedure as \textbf{self-involvement} in this paper. 
Theoretically speaking, this procedure would give the pre-trained model a constrained semantic sample space for learning. Thus, the model would know what they need to learn.
At inference stage, we assume relevance scores $o$ can reflect the difficulty (or quality) of negative samples. We select $K_i$ hardest negative samples from negative samples (normally, $K_i \ll N$) at $i$-th compressor and feed them to the next compressors:
\begin{align}
    \begin{split}
        S_{i+1}  &= \text{Top}( f_{i} (S_i), K_i).  \\
    \end{split}
\end{align}
Note that the value of K is a hyper parameter and will gradually decrease. The positive sample will automatically send to the next compressor. The last compressor only need to classify the samples, we regard it as the classifier in SIR. In strategy V1, every compressor is initialized at the same checkpoint of certain pre-trained model.

\subsection{Strategy V2: Self-involvement with Hard Samples and Parameters Memory}

For faster convergence, in strategy V2, we keep the same network architecture as strategy V1 but let the model parameters to flow from one compressor to the next. After having trained the front compressor, we share the front compressor's parameter with the rear one instead of training it from ground up. The rear compressor would start at the checkpoint that the front compressor has achieved. This is a soft parameter sharing mechanism~\cite{ruder2017overview}.

The reason we use the soft parameter sharing mechanism is that after training the previous compressor $C_{i-1}$, 
it would capture the complete distribution information of the previous training samples. Sharing parameters can be interpreted as adding prior knowledge to the next compressor $C_i$. 
For compressor $C_i$, the semantic space of negatives would be refined based on both previous memorized knowledge and current training samples.
So it would have a more comprehensive sample distribution compared with strategy V1. 



\subsection{Strategy V3: Self-involvement with Dynamic and Hard Samples}
Directly employ the prior knowledge in model initialization would improve model performance to some extent, but the distribution of negative samples used for training each compressor is still fixed. The fix distribution of negative samples leads to the dynamics loss, and that can not be compensated by parameter memory.
Therefore, V3 uses the supervisory signal to synchronously adjust the compressors at all levels, so that the negative sample distribution selected by each compressor is also dynamic. V3 adopts hard parameter sharing mechanism~\cite{ruder2017overview}.

In the forward process of the training stage, we keep the same inference procedure as the previous two strategies. Differently, in the backward process, when the model parameters are updated with the supervisory signal, the negative sample distribution selected by the front compressor for the rear compressor will also be adjusted. With the continuous convergence of the model, the difficulty of negative samples increases gradually, which is consistent with the principle of curriculum learning.

Conclusively, We iteratively select dynamic hard samples and gradually force the train set updating step by step. For query $q$, the loss of the whole model consists of classification loss of the first compressor and the last compressor:
\begin{align} \label{V3loss}
    \begin{split}
        \mathcal{L} ( S(q), \theta ) = \mathcal{L}(S(q), \theta, 1) + \mathcal{L}(S(q), \theta, M).
    \end{split}
\end{align}

Since the compressors in strategy V3 share their parameters, we can formulate the learning process as:
\begin{align} \label{V3theta}
    \begin{split}
    \theta = \mathop{\arg\min_{\theta}}\sum_{q\in Q} \mathcal{L} (S(q), \theta).
    \end{split}
\end{align}

\begin{algorithm}[htb]
\caption{ Strategy V3 \& V4 }
\label{alg:Framework}
\begin{algorithmic}[1] 
\REQUIRE ~~\\ 
    Pre-trained model $f$; compressor $C$ parameterized by $\theta$;\\
    Input data $S_1$; the number of selected negative samples at $i$-th compressor $K_i$.\\
    \renewcommand{\algorithmicrequire}{ \textbf{Initialization:}}
    \REQUIRE Initializing compressor $C$ with $f$.\\
    \renewcommand{\algorithmicrequire}{ \textbf{Loop:}}
    \REQUIRE for i from compressor 1 to compressor M~~\\
    \STATE Forwarding $C$ based on $S_i$;
    \STATE Selecting $K_i$ top ranked negative samples and output $S_{i+1}$;
    \renewcommand{\algorithmicrequire}{ \textbf{Do:}}
    \REQUIRE 
    \STATE Calculating loss with \eqref{V3loss} for strategy V3, \eqref{v4loss} for strategy V4;
    \STATE Updating $\theta$ with \eqref{V3theta};
    \renewcommand{\algorithmicrequire}{ \textbf{Output:}}
    \REQUIRE Fine-tuned model $C$. 
\end{algorithmic}
\end{algorithm}



\subsection{Strategy V4: Self-involvement with Controlled Dynamic and Hard Samples} 


In strategy V3, the negative sample distribution generated by compressor at each level is adjusted with supervisory signal. However, the signal of the following stage compressor can not directly feed back to supervise the previous stage compressor, but can only indirectly rely on the parameter sharing mechanism to affect previous compressor.



In order to solve the problem of the front compressor's weak perception of the ranking results of rear compressor, we use the feedback mechanism of the control theory~\cite{franklin2002feedback} for reference and use the conditional probability (CPR) to link the ranking results of the front and rear compressors. Then, the supervisory signal can be directly back-propagated from the rear compressor to its front compressors.


In the modeling of V4, for query $q$, $i$-th compressor outputs is samples' relevance score $o_i \in R^{1+|S_i|}$. We softmax $o_i$ as in Eq. (\ref{softmax}) to obtain $P(o_i)$. We denote $P(o_i)$ as $P_i$ for simplicity.

Here, we constrain the relative position of same top-ranked samples in different level of compressors, which means that for $j$-th sample at $i$-th compressor, its predicted score at $i$-th compressor is $o_i^j$ and its score at front compressor is $o_{i-1}^j$. Note that the positive sample's score is $o_{i}^0$.

We denote the samples' conditional probability at $i$-th compressor as $\text{CPR}_i \in R^{1+|S_i|}$. $\text{CPR}_i$ is computed by the multiplication of posterior probability of sample to fulfill the goal that the supervisory signal of each compressor can be back propagated to front compressor directly. Such settings can control sample dynamics at front compressor with supervisory signal at rear compressor, thus achieve better semantic space of negative samples at all compressors. The conditional probability of $i$-th compressor can be written as:

\begin{align}
    \begin{split}
        \text{CPR}_i = \textbf{softmax} (P_1^{0 : |S_i|} \odot  P_2^{0 : |S_i|} \odot ... \odot P_i^{0 : |S_i|}), 
    \end{split}
\end{align}
 where $\odot$ denotes element wise multiplication and $P_i^{0:|S_i|}$ denotes the beginning $1+|S_i|$ logits of $P_i$.

Then, we formulate the loss function based on $i$-th compressor for query $q$ as:
\begin{align}
\begin{split}
        \mathcal{L}(S(q), \theta, i) =  - \log \text{CPR}(o^{0}_i) - \sum_{j=1}^{|S_i|}\log (1- \text{CPR}( o^{j}_i )). 
\end{split}  
\end{align}
The loss function of V4 is:
\begin{align}
\begin{split}
        \mathcal{L}(S(q), \theta) =  \sum_{i=1}^M L(S(q),\theta, i).    \label{v4loss}
\end{split}  
\end{align}

Pseudo code for strategy V3 and V4 can be found in Algorithm \ref{alg:Framework}.

\section{Experiments and Analysis} \label{Experiment}
\subsection{Datasets}

\textbf{MS MARCO Document Ranking} is now the most popular dataset in IR~\cite{bajaj2016ms}. There are 3.2 million documents and 367,013 queries in the training set and the goal is to rank based on their relevance.

\subsection{Pre-train Models}
We conduct experiments on 2 pre-trained language models and 3 pre-trained IR models. For pre-trained language models, BERT~\cite{devlin2019bert} and ELECTRA~\cite{clark2020electra} are used. BERT designs Masked Language Models (MLM) and Next Sentence Prediction (NSP) as its pre-training tasks. Instead, ELECTRA uses replaced token detection task. For pre-trained IR models, WLP uses Wiki Link Prediction task, ICT uses Inverse Cloze task~\cite{DBLP:journals/corr/abs-1906-00300,DBLP:journals/corr/abs-2002-03932} and PROP uses Representative Words Prediction task~\cite{10.1145/3437963.3441777}. 


\subsection{Dataset Pre-processing}
For each positive sample, we first find its corresponding negative samples with HDCT and ANCE.

\subsubsection{HDCT} is a context-aware document term weighing framework for document indexing and retrieval~\cite{dai2019contextaware}. It first uses passage level term weighting and then aggregates them into the document level term weighting. Adopting HDCT indexing can effectively improve the retrieval accuracy for various fine-tuned pre-trained models.

\subsubsection{ANCE} is a learning mechanism which use a asynchronously updated ANN index to select hard negative samples from the whole corpus to construct datasets for pre-train models~\cite{xiong2020approximate}. It can achieve state-of-the-art performance and maintain efficiency at the same time.

\subsection{Implementation Details}

Considering the computational cost, we used three-level compressor (including classifier) for the sample selection. Increasing number of level could require a large amount of data for training. In our implementation, we found that three-level architecture would achieve SOTA compared with other multiple level architectures. To show that SIR can be implemented on various pre-trained models, \textbf{we use pre-trained BERT-base and ELECTRA-base with limited ranking ability}.

For training setup, considering that distributed data parallel is used for experiments acceleration, we set samples size to be the integer times of the number of GPU devices. Since we train our model with V100 which has 8 GPUs, we set the size of a training block (the sample number of positives and negatives in a block) as 88 and batch size of the training block as 4. We set our self-involvement samples compressed from 88 to 48 and 48 to 16 in 3 level models. And we further compress 16 to 8 in 4 level models. 
Also, we set the training epochs as 2. In this work, we used AdamW as our Optimizer with $\beta_1 = 0.9$ and $\beta_2=0.999$, and learning rate as $5e^{-5}$. We adopt both the ANCE based top 100 dataset and HDCT based top 100 dataset as the test bed. 
Since we are trying to leverage pre-trained models with simple SIR strategies, here we only train our model on the side of fine-tuning downstream ranking tasks based on top 100 dataset.

\subsection{Evaluation}
We evaluate SIR on ranking tasks and use mean reciprocal rank (MRR), mean average precision (MAP) and normalized discounted cumulative Gain (NDCG) as metrics.

\subsection{Performance Comparison}
In Table \ref{IR table}, we show that SIR can substantially enhance pre-trained IR models. WLP~\cite{DBLP:journals/corr/abs-2002-03932}, ICT~\cite{DBLP:journals/corr/abs-2002-03932} and PROP~\cite{10.1145/3437963.3441777} are three pre-trained IR models with different pre-training tasks. All of them gain great performance improvement on both HDCT top 100 based data and ANCE top 100 based data. Only V4 results are reported in Table \ref{IR table}.

\begin{figure}[t]
     \centering
         \includegraphics[width=0.33\textwidth]{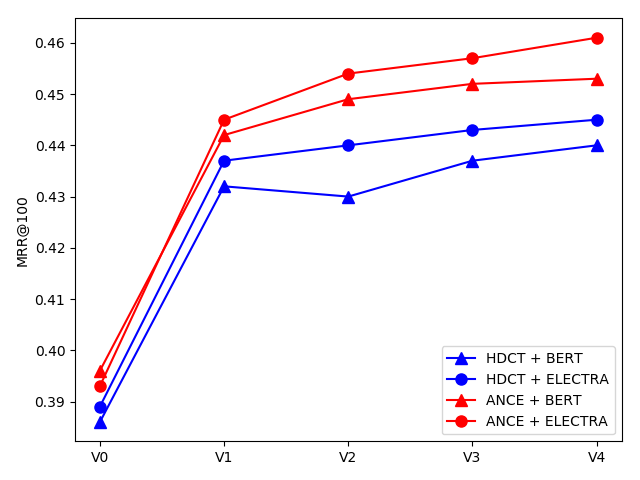}
         \caption{Different Strategies. V0 represents pre-trained model fine-tuned with random-selected negative samples.}
         \label{SIRcompare}
\end{figure}

In Table \ref{table2}, we compare different SIR strategies based on two widely used pre-trained language models, including BERT-base and ELECTRA-base. Results show that all SIR strategies increase models' performance. Among 4 strategies, V4 is the best and V3 is close to V4. It empirically shows that dynamic and hard negative samples would be beneficial for fine-tuning of the pre-trained models. The results are also shown in Figure \ref{SIRcompare}.

To conclude, experiments show that SIR can reach SOTA ranking performance even under a native BERT base and it can also greatly improve arbitrary pre-trained model's performance on MS MARCO Document ranking. \textbf{We used SIR V4 strategy to boost ELECTRA-large, which achieved the best scores in MSMARCO document ranking leaderboard in May 2021. In order to prevent exposure of information, the leaderboard scores are not provided in tables for the time being}.

\begin{figure}[t]
     \centering
         \includegraphics[width=0.33\textwidth]{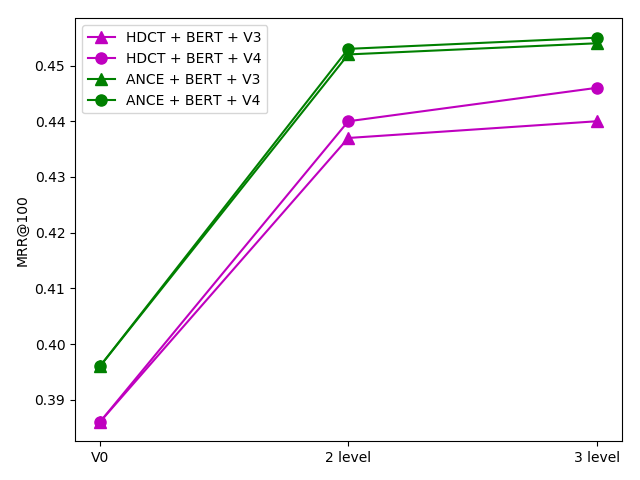}
         \caption{Different Number of Level. V0 represents pre-trained model fine-tuned with random-selected negative samples.}
         \label{multilevel}
\end{figure}

\begin{table*}[]
\centering
\begin{tabular}{lllllll}
\toprule
\multirow{2}{*}{Model} & \multicolumn{3}{c}{HDCT Top 100} & \multicolumn{3}{c}{ANCE Top 100} \\
           & MRR@100 &MAP@20 &NDCG@20 & MRR@100 &MAP@20 &NDCG@20\\ \midrule

WLP       & 0.361 & 0.358 & 0.446 & 0.374 & 0.370 & 0.464 \\
ICT       & 0.385 & 0.383 & 0.470 & 0.394 & 0.391 & 0.484 \\
PROP      & 0.391 & 0.388 & 0.474 & 0.397 & 0.394 & 0.486 \\
\midrule
WLP + SIR   & \textbf{0.435}\dag & \textbf{0.434}\dag & \textbf{0.517}\dag & \textbf{0.449}\dag & \textbf{0.446}\dag & \textbf{0.535}\dag \\
ICT + SIR   & \textbf{0.440}\dag & \textbf{0.438}\dag & \textbf{0.521}\dag & \textbf{0.457}\dag & \textbf{0.455}\dag & \textbf{0.543}\dag \\
PROP + SIR  & \textbf{0.443}\dag & \textbf{0.441}\dag & \textbf{0.523}\dag & \textbf{0.454}\dag & \textbf{0.451}\dag & \textbf{0.540}\dag \\
\bottomrule

\end{tabular}
\caption{Comparison between pre-trained IR models with the combination of SIR V4, \dag denotes our results (SIR) are statistically significant.} \label{IR table}
\end{table*}

\begin{table}[]
\setlength{\tabcolsep}{2pt}
\begin{tabular}{@{}lllll@{}}
\toprule 
\multirow{2}{*}{Model} & \multicolumn{2}{c}{HDCT Top 100} & \multicolumn{2}{c}{ANCE Top 100} \\
              & \small{MRR@100} & \small{MRR@10} & \small{MRR@100} & \small{MRR@10} \\ \midrule
BERT          & 0.386  & 0.377 & 0.396   & 0.386  \\\midrule
BERT + V1 & 0.432\dag  & 0.425\dag & 0.442\dag   & 0.434\dag  \\
BERT + V2 & 0.430\dag  & 0.423\dag & 0.449\dag   & 0.441\dag  \\
BERT + V3 & \textbf{0.437}\dag  & \textbf{0.430}\dag & \textbf{0.452}\dag   & \textbf{0.444}\dag  \\
BERT + V4 & \textbf{0.440}\dag   & \textbf{0.433}\dag  & \textbf{0.453}\dag   & \textbf{0.445}\dag       \\\midrule


ELECTRA       & 0.389  & 0.380 & 0.393   & 0.383  \\\midrule
ELECTRA+V1  & 0.437\dag  & 0.430\dag & 0.445\dag   & 0.437\dag  \\
ELECTRA+V2  & 0.440\dag  & 0.433\dag  & 0.454\dag   & 0.446\dag  \\
ELECTRA+V3  & \textbf{0.443}\dag  & \textbf{0.437}\dag & \textbf{0.457}\dag   & \textbf{0.450}\dag  \\
ELECTRA+V4  & \textbf{0.445}\dag   & \textbf{0.438}\dag  & \textbf{0.463}\dag   & \textbf{0.455}\dag   \\ \bottomrule
\end{tabular}
\caption{Comparison between native BERT with our fine-tuning strategies, \dag denotes our results (SIR) are statistically significant. The result of SIR V1 and V2 are presented under the random sample selection while V3 and V4 are presented under the top K sample selection.} \label{table2}
\end{table}

\begin{table}[]
\setlength{\tabcolsep}{2pt}
\begin{tabular}{@{}lllll@{}}
\toprule
\multirow{2}{*}{Model} & \multicolumn{2}{c}{HDCT Top 100} & \multicolumn{2}{c}{ANCE Top 100} \\
                    & \small{MRR@100} & \small{MRR@10} & \small{MRR@100} & \small{MRR@10} \\ \midrule
BERT          & 0.386  & 0.377 & 0.396   & 0.386  \\\midrule
V1 Random 16     & 0.432\dag  & 0.425\dag & 0.442\dag   & 0.434\dag  \\
V1 Top 16        & 0.319\dag  & 0.308\dag & 0.320\dag   & 0.308\dag \\ \midrule
V2 Random 16     & 0.430\dag  & 0.423\dag  & 0.449\dag   & 0.441\dag \\
V2 Top 16        & 0.324\dag  & 0.312\dag &  0.328\dag  & 0.315\dag \\\midrule
V3 Top 16        & \textbf{0.437}\dag  & \textbf{0.430}\dag & \textbf{0.452}\dag   & \textbf{0.444}\dag \\
V3 Random 16     & \textbf{0.436}\dag  & \textbf{0.429}\dag & \textbf{0.450}\dag   & \textbf{0.442}\dag \\
V3 Top 8         & \textbf{0.440}\dag  & \textbf{0.434}\dag & \textbf{0.454}\dag   & \textbf{0.446}\dag\\
V3 Random 8      & \textbf{0.435}\dag  & \textbf{0.429}\dag & \textbf{0.453}\dag   & \textbf{0.445}\dag \\\midrule
V4 Top 16        & \textbf{0.440}\dag  & \textbf{0.433}\dag & \textbf{0.453}\dag   & \textbf{0.445}\dag \\ 
V4 Random 16     & \textbf{0.436}\dag  & \textbf{0.429}\dag & \textbf{0.454}\dag   & \textbf{0.446}\dag \\
V4 Top 8         & \textbf{0.446}\dag  & \textbf{0.440}\dag & \textbf{0.455}\dag   & \textbf{0.447}\dag       \\
V4 Random 8      & \textbf{0.442}\dag  & \textbf{0.435}\dag & \textbf{0.452}\dag   & \textbf{0.444}\dag    \\ \bottomrule
\end{tabular}
\caption{Comparison between BERT with different fine-tuning strategies. \dag denotes our results (SIR) are statistically significant.} \label{table3}
\end{table}

\subsubsection{Dynamic VS Randomness}
In our experiments, we find that introducing randomness in training would have almost the same effect as using model dynamic. We set up several competitive experiment on our four strategies. In some settings, neural model-based compressor is replaced by random compressor to capture the distribution of whole training samples. Instead of selecting K samples with the highest scores (marked as Top in Table \ref{table3}), random compressor select K samples randomly (marked as Random in Table \ref{table3}).

As we can observe from Table \ref{table3}, when we directly use top 16 hard negative samples in each training step on strategy V2, the performance would have a severe decline. Surprisingly, using random 16 samples for training instead of top 16, the performance of strategy V2 significantly improved. We think that when dynamic negative samples are not selected by the model, the randomness of the data can provide a certain dynamics to improve the effect. However, when we introduce randomness to V3 and V4 strategies, which select dynamic and hard negative samples, the performance would not have an obvious improve. \textbf{This shows that the randomness of data has a certain homogeneity with the dynamics based on the model}.


\subsubsection{Impact of Multi-level compressors structure}
In Table \ref{table3}, we explore the impact of number of levels for V3 and V4. Results are also shown in Figure \ref{multilevel}. For 3-level compressors structure (including classifier), model selects samples from 88 to 48 to 16 for updating. And for 4-level compressors structure, model selects samples from 88 to 48 to 16 to 8, with an additional compressor. In Table \ref{table3}, for both Random setting and Top setting, the performance is not necessarily improved as the number of levels increases. This may because with the decreasing of negative samples in training, although hard negative samples would be the most informative ones, they lack the ability of capturing full distribution of the training samples. Thus, with more levels of compressors, the performance could not be improved significantly.

\begin{figure}[]
     \centering
     \begin{subfigure}[b]{0.14\textwidth} \label{V1S}
         \centering
         \includegraphics[width=\textwidth]{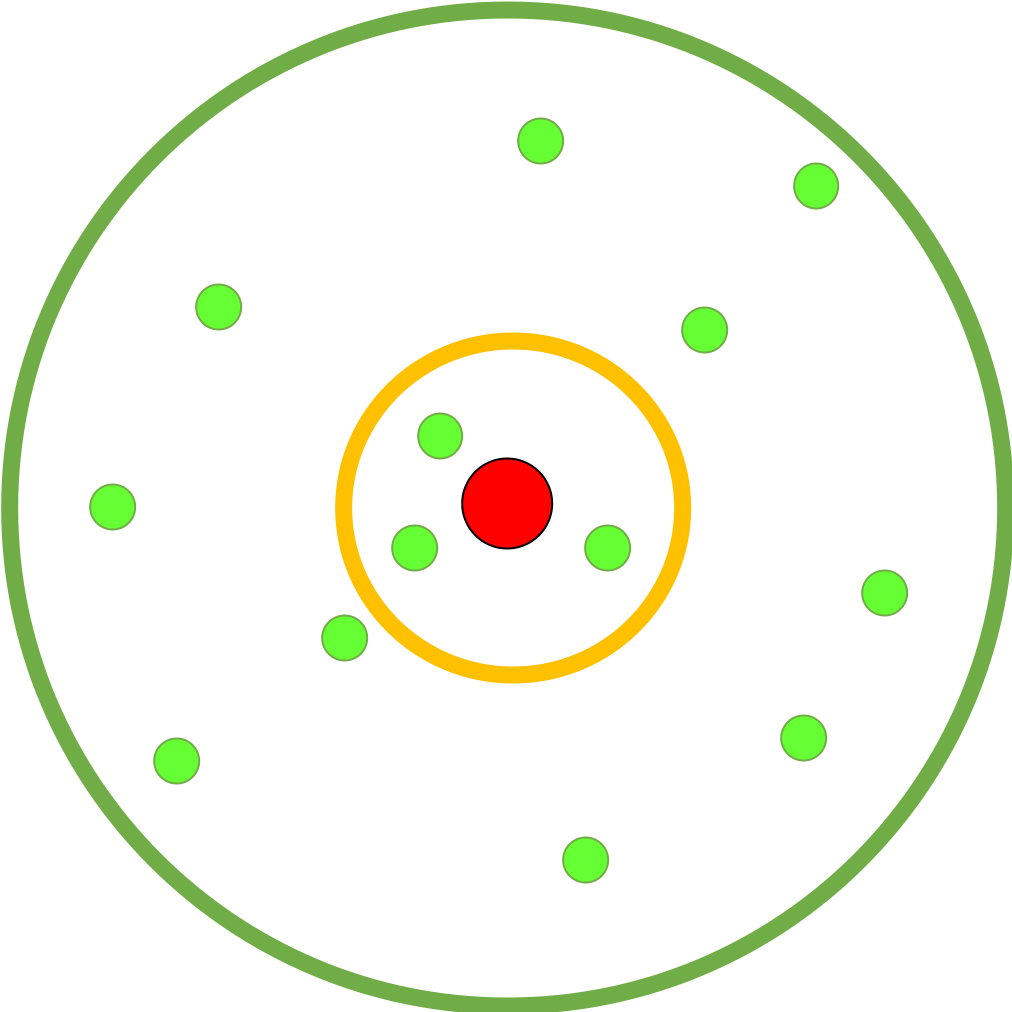}
         \caption{V1}
         \label{V1S}
         
     \end{subfigure}
     \hfill
     \begin{subfigure}[b]{0.14\textwidth}
         \centering
         \includegraphics[width=\textwidth]{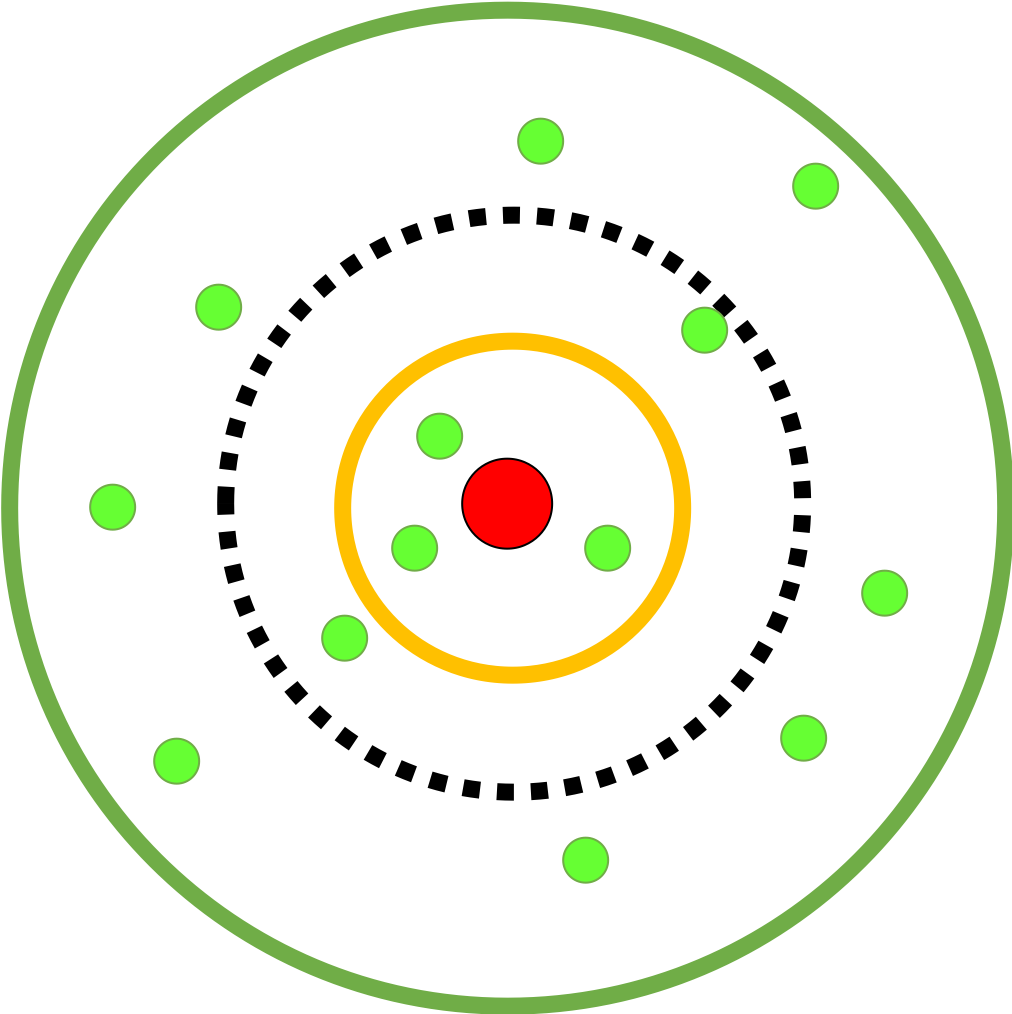}
         \caption{V2}
        \label{V2S}
     \end{subfigure}
     \hfill
     \begin{subfigure}[b]{0.14\textwidth}
         \centering
         \includegraphics[width=\textwidth]{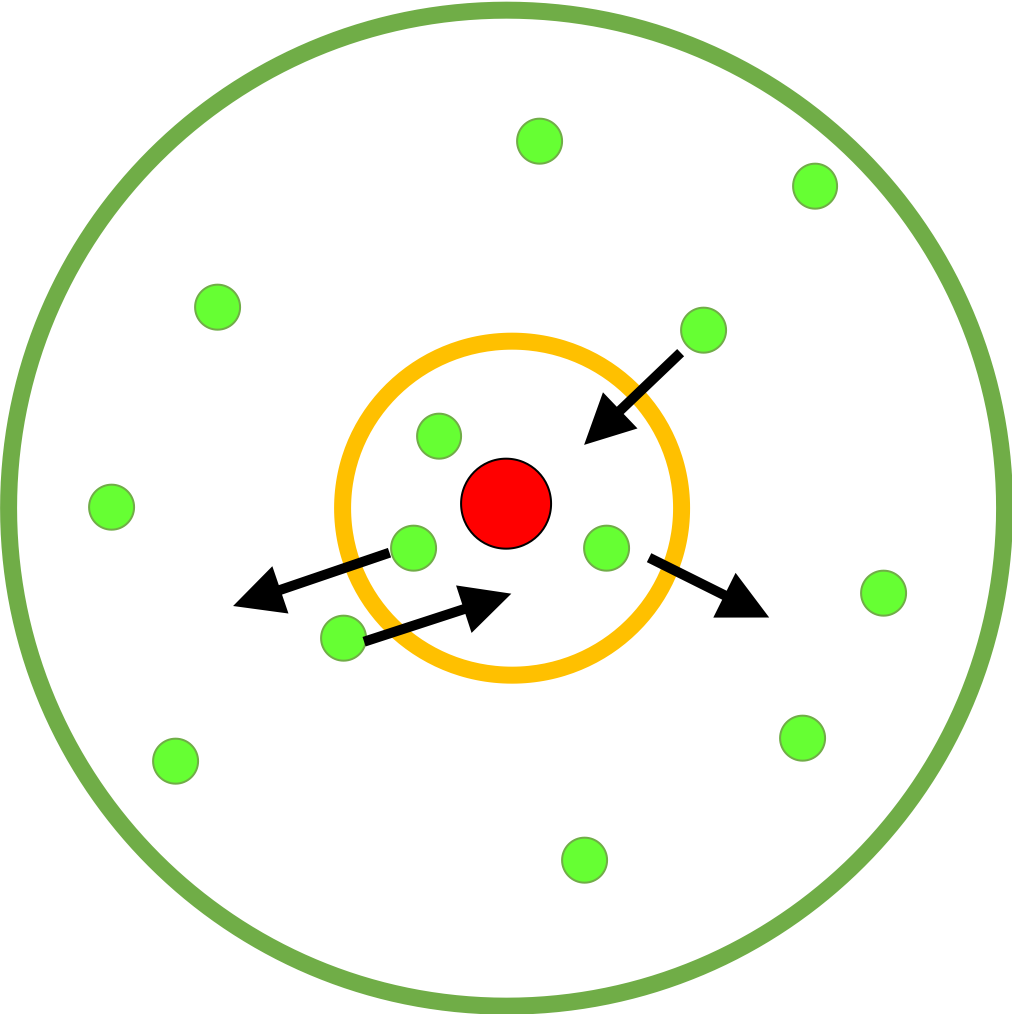}
         \caption{V3\&V4}
         \label{V3V4S}
    \end{subfigure}
        \caption{Semantic Space of Negatives in Classifier.}
        \label{SIRspace}
\end{figure}

\subsection{Strategy Analysis}
Figure \ref{SIRspace} shows semantic space of negatives in classifier. Markers in the figure have the same meaning as Figure \ref{fig2}. Note that Figure \ref{fig2} shows semantic space of the whole model, while Figure \ref{SIRspace} shows semantic space in classifier. V1 selects fix hard negative samples as in Figure \ref{V1S}. V2 selects fix hard negative samples but with distribution information of wider space as in Figure \ref{V2S}, where black dash line represents distribution information received from front compressor. V3 and V4 selects dynamic hard negative samples as in Figure \ref{V3V4S}. The difficulty of negative samples changes with the training process. The comparison of the four strategies (V1-V4 of SIR) are shown in Table \ref{table2}.

\subsubsection{Analysis to strategy V1}
Strategy V1 directly uses multi-level model to select hard negative samples for fine-tuning pre-trained models. 
Negative samples can maintain a relatively normal effect only under the little dynamics brought by random data sampling. Once this dynamics is lost, the effect will decrease a lot, which has been shown in Table \ref{table3}. Since the compressors at each level of V1 are fixed, the semantic space of negative samples is not improved or gradually narrowed, resulting in poor generalization ability.


\subsubsection{Analysis to strategy V2}
Strategy V2, which adopts a soft parameter sharing mechanism, has a certain improvement over V1. Such results show that the prior knowledge from parameters memory would lead models to learn in a more comprehensive manner. Compared with strategy V1, V2 can improve the narrow semantic space with the prior knowledge to prevent the model getting trapped into local optimal.

\subsubsection{Analysis to strategy V3}
Strategy V2 proves the effectiveness of soft parameter sharing. Strategy V3 goes further to use the supervisory signal to adjust the distribution of negative samples.
Compared with V1, V2 and baseline pre-trained models, strategy V3 outperforms it a lot. With the guidance of supervisory signal, V3 would re-evaluate the difficulty of negative samples after parameter update. Therefore, the model can obtain the most suitable negative samples at any stage, which means that the dynamic distribution of negative samples is learned.

\subsubsection{Analysis to Strategy V4}

V3 dynamically adjusts the distribution of negative samples by using the supervisory signal through the parameter sharing of the compressor. However, the rear compressor cannot feed back the rationality of negative sample selection to the front compressor. In V4, the feedback loop of the front and rear stage compressor is formed through conditional probability, so that the supervisory signal can be adjusted directly through the feedback loop to control the dynamics brought by parameter sharing. Under a certain dynamics control, V4 outperforms other strategies. The negative sample semantic space of V4 is the same as that of V3. The only difference is that supervisory signals from upper compressors can also refine lower compressors, thus they can influence semantic space in lower compressors.

\section{Conclusions and Future Work} \label{Conclusion}
In this paper, we propose SIR, a light-weight fine-tuning strategy for pre-trained models. The key idea is to optimize semantic space of negatives for fine-tuning ranking models. SIR can enhance various pre-trained models without complex design in tasks or additional computational resources. Experiments on MS MARCO document ranking dataset shows that SIR achieves significant improvement over the baseline.

Currently, compressors are all homogeneous. We think it promising to explore heterogeneous compressors to integrate strength of different pre-trained models in the future.

\bibliography{aaai22}

\clearpage

\appendix

\end{document}